\documentclass[journal]{IEEEtran}

\usepackage{cite}   
\usepackage{orcidlink}
\usepackage{amsmath,amssymb,amsfonts}
\usepackage{dsfont} 
\newtheorem{proposition}{Proposition}
\usepackage{algorithm}
\usepackage{algpseudocode}
\usepackage{graphicx}
\usepackage{textcomp}
\usepackage{xcolor}
\usepackage{array}
\usepackage{multirow}
\usepackage{booktabs}
\usepackage{url}

\graphicspath{{../Iagaes/}} 

\newcommand{\Npwr}{N_{\mathrm{pwr}}}
\newcommand{\Nsig}{N_{\mathrm{sig}}}
\newcommand{\Gsig}{G_{\mathrm{sig}}}
\newcommand{\Sdiag}{S_{\mathrm{diag}}}
\newcommand{\Ereach}{E_{\mathrm{reach}}}
\colorlet{red}{black}
\begin{document}

\title{An LLM-Orchestrated Neuro-Symbolic Framework for Verifying Diagnostic Signal Reachability in Electronic Systems}

\author{Heba~Jaradat~\orcidlink{0000-0003-0698-7427},
        Mohammed-Khalil~Ghali ~\orcidlink{0009-0003-5109-7084},
        Saurabh~Kulkarni ~\orcidlink{0009-0009-6720-1126},
        Rohan~Kulkarni ~\orcidlink{0009-0007-1398-4327},
        and~Sangwon~Yoon ~\orcidlink{0000-0002-1613-0745}
\thanks{(\textit{Corresponding author: Sangwon Yoon.})}
\thanks{Heba Jaradat, Mohammed-Khalil Ghali, and Sangwon Yoon are with the School of Systems
Science and Industrial Engineering, Binghamton University, State University of
New York, Binghamton, NY 13902 USA (e-mail: hjaradat@binghamton.edu;
mghali1@binghamton.edu; yoons@binghamton.edu).}
\thanks{Saurabh Kulkarni and Rohan Kulkarni are with Meta Platforms, Inc., 1 Hacker Way,
Menlo Park, CA 94025 USA (e-mail: saurabhkulkarni@meta.com; irohan@meta.com).}}

\markboth{}%
{Jaradat \MakeLowercase{\textit{et al.}}: Neuro-Symbolic Verification of Diagnostic Signal Reachability}

\maketitle

\begin{abstract}
Diagnostic Signal Reachability (DSR) captures whether board-level health signals
are observable by management controllers, including baseboard management
controllers (BMCs), complex programmable logic devices (CPLDs), and other smart
controllers. Verifying DSR is essential for field diagnosability in datacenter
fleets, where one design is deployed at very large scale, yet it remains
largely manual, since each signal is traced by hand across schematic sheets,
renamed nets, series components, and monitoring devices. Large language models
(LLMs) are attractive but unreliable at structural connectivity reasoning,
especially for renamed or indirectly observed signals, where over-claiming can
silently hide a rare coverage gap. We present NetFlow-DSR, a neuro-symbolic
framework in which an LLM orchestrates the end-to-end DSR pipeline and resolves
device roles mainly from part-name families and library pin names, consulting
device specifications only for the few parts that remain ambiguous. Pin
directionality is read directly from the design library, and deterministic graph
engines verify
connectivity over the authoritative netlist. The pipeline detects
diagnostic nets, canonicalizes renamed net fragments, and performs
controller-seeded, power-aware breadth-first search over an admissibility graph
that credits device-mediated monitoring. Each verdict carries an auditable
per-signal trace card whose proof path is the witness. Across five production
boards from an AI platform, covering 421 expert-graded in-scope signals, a
three-stage ablation improves pooled F1 from 0.578 to 0.944. Because a monitored
verdict is emitted only with such a proof path, the verifier is sound by
construction, and precision and specificity were 100\% in the evaluated data. In
a cross-model reproducibility study \textcolor{black}{on the system-controller module, precision}
remained 100\% across three LLMs while recall tracked each model's
device-semantic extraction. Precision therefore depends on the netlist-grounded verifier and correct device semantics rather than on any single LLM.
\end{abstract}

\begin{IEEEkeywords}
Diagnostic coverage, Diagnostic signal reachability, Electronic design automation, Generative AI, Hardware verification, Large language models,
Netlist analysis, Neuro-symbolic verification.
\end{IEEEkeywords}

\section{Introduction}
\IEEEPARstart{M}{odern} AI hardware servers  and datacenter platforms expose hundreds of
diagnostic signals, including power-good, presence, reset, interrupt, alert, I2C,
and ADC telemetry nets. Fault detection and field recovery depend on these
signals being observable by a management controller, typically a BMC or CPLD. If a diagnostic signal is not observable by the
management path, the associated failure mode may remain invisible, increasing
no-fault-found returns, service time, and system downtime. The cost is highest at
datacenter scale. A hyperscale fleet deploys the same board design in very large
numbers, so a single unobservable signal in one design is copied into every
deployed unit. In large AI training clusters, one failing node can interrupt a job
that spans thousands of accelerators, and fast fault isolation depends on the
management path seeing the failure. A coverage gap found after deployment can
require a board respin or a fleet-wide workaround, whereas the same gap found
during design review costs only a design change. DSR is therefore a
serviceability-critical sign-off activity, typically performed during
pre-production design validation. Before deployment, engineers confirm that
faults on power rails, devices, connectors, and telemetry paths can be observed
by management firmware. In current practice, however, DSR is verified
manually, by first identifying the diagnostic signals and then tracing each one
across schematic sheets and through intermediate components. This process is
slow. For complex modules such as motherboards (MBs) and accelerators, reporting
the diagnostic signals and then verifying each one across the schematics can
require several weeks per board. It is also difficult to audit and increasingly
error-prone as board complexity grows. The central challenge is to detect the
in-scope diagnostic signals and verify, exhaustively and reproducibly, whether
each one reaches an appropriate management controller.

What makes this verification difficult is that the electrical evidence for a
single signal is rarely local to one net or one schematic sheet. A physical
signal is frequently split into several differently named fragments as it crosses
a series resistor, ferrite bead, or zero-ohm link, so the fragment a controller
actually touches may carry a different name from the one an engineer happens to
inspect. Observability is also frequently indirect. A rail or pin may never
connect electrically to a controller yet still be observed by a power monitor, an
analog-to-digital converter, a temperature sensor, or a GPIO expander that relays
the measurement or pin state to firmware over I2C. Legitimate paths further pass
through devices such as I2C multiplexers, bus buffers and repeaters, and level
translators, whereas high-fan-out power rails and grounds are better treated as
barriers, because tracing through them would make almost every signal appear
connected to every controller. A reviewer therefore has to hold signal identity,
admissible device hops, and rail exclusion in mind at once while following each
signal across many sheets, a cognitive burden that scales poorly as board
complexity grows. The two possible mistakes are, moreover, not equally costly.
Reporting a genuinely unobservable signal as monitored silently removes a real
coverage gap from the review set and lets a fault path escape scrutiny, whereas
the opposite error merely returns an already-covered signal to a bounded,
self-correcting review. Two requirements therefore follow for an automated
verifier. It should be deliberately biased against over-claiming and make every
positive verdict auditable rather than merely plausible.

These two requirements give a useful basis for weighing the automation options
available. Learning-based methods are a natural first candidate, and they have
advanced rapidly across electronic design automation (EDA). Machine learning has
been applied to EDA for over a decade \cite{huang2021machine}, and large language
models have since been used for HDL generation and debugging
\cite{thakur2024verigen,pinckney2025revisiting,qayyum2025llm}, EDA-workflow
orchestration \cite{wu2024chateda}, analog design automation and sizing
\cite{liu2025eesizer,wei2026toposizing}, analog-netlist comprehension
\cite{duaescu2025assessing}, and netlist generation from natural-language
descriptions \cite{ye2025llm4netlist}. These systems synthesize, orchestrate,
size, or generate design artifacts. None, however, verifies whether board-level
diagnostic signals reach a management controller.

There is also reason to be careful about delegating the structural part of the
problem to a learned model. Neural executors of classical graph algorithms
degrade out of distribution \cite{velivckovic2022clrs}, and vision-language
models read graph nodes far more reliably than edges when structure is extracted
from images \cite{zhao2026vision} and fine-grained visual understanding remains an active architectural problem in its own right\cite{11382543}; large language models reasoning over
linearized graphs have likewise been reported to assert or deny edges that are
explicitly encoded in the input, a failure termed structural hallucination
\cite{wang2026large}. Taken together, these findings suggest that structural
reasoning is better grounded in an authoritative graph than inferred from images,
text, or net names.

Such a graph can be searched with primitives that are individually well
established. Renamed fragments of the same physical signal can be reconciled as
an equivalence problem and evaluated efficiently with disjoint-set union
\cite{galler1964improved,tarjan1975efficiency}. Controller observability is a
reachability problem solved by breadth-first search with a parent tree that
supplies a witness path \cite{moore1959shortest,cormen2022introduction}. Deterministic graph search underpins connectivity analysis
  in design automation and in reliability engineering alike,
  in global routing
\cite{huang2021detailed}, in signal-flow-based failure-mode
  analysis\cite{5221000}, and efficient reachability querying is itself a
well-studied problem \cite{cohen2003reachability,yu2010graph}. These primitives
are standard. Applying them directly is nonetheless insufficient for DSR, because
the verifier still has to define which renamed nets are identical, which device
hops are admissible, which rails act as non-conducting walls, and which
functional observations through monitor devices count as coverage.

These observations point to a division of the work rather than to a choice
between the two families. NetFlow-DSR, the netlist-grounded neuro-symbolic
framework presented here, separates learned semantics from computed structure.
Instead of perceiving connectivity from schematic images or from net names, it
computes connectivity deterministically over the authoritative electrical graph
that schematic-capture tools already emit. A large language model orchestrates
the pipeline and supplies only the semantics that a netlist cannot encode. It
identifies the management controllers, selects the in-scope diagnostic signals,
and resolves device roles mainly from part-name families and pin names, consulting
device specifications only where these do not suffice. Pin names and
directionality are read directly from the design's own EDA library. Two classical solvers then carry the structural
burden. Disjoint-set Union-Find reconciles renamed fragments into canonical
signals \cite{galler1964improved,tarjan1975efficiency}, and controller-seeded,
power-aware breadth-first search decides reachability and returns a parent tree
that serves as a witness path \cite{moore1959shortest,cormen2022introduction}.
A monitored verdict is emitted only when this procedure exhibits an explicit,
netlist-verified path to a controller, recorded as a per-signal trace card. The
verifier is therefore sound by construction with respect to its admissibility
model, and interpretable by design rather than through a post hoc rationale.

Because NetFlow-DSR reports coverage figures, a terminological clarification is required before the formulation is
  presented. We use ``diagnostic coverage'' throughout in the
  observability sense, that is, the fraction of in-scope diagnostic
  signals for which an admissible management path to a controller is
  proven. This is distinct from the failure-rate-weighted diagnostic
  coverage defined in functional-safety standards such as IEC~61508 and
  ISO~26262, in which detected failure modes are weighted by their
  contribution to the dangerous failure rate. The two are related as
  precondition and realization. A failure mode whose diagnostic signal
  never reaches a management controller cannot be detected by management
  firmware, irrespective of the detection logic that firmware implements,
  so electrical reachability is a necessary condition for detection through the management path. NetFlow-DSR verifies this
  precondition exhaustively and emits the per-signal observability verdict
  that a weighted computation consumes as input. It does not compute the
  weighting itself, which requires component-level failure-rate data and a
  mapping of signals to failure modes that lie outside the scope of this
  work. The metrics reported in Section~IV are accordingly verification-level
  quantities over the expert-labeled signal set, not certified coverage figures
  under a functional-safety standard.

This paper makes the following contributions.
\begin{enumerate}
\item Ground connectivity in the netlist's exact edges instead of perceiving it
from images, removing the edge-recovery weakness (direction and polarity) that
dominates vision-graph methods. Pin direction is read from the design's EDA
library, polarity is decoupled from the observability verdict, and the LLM
supplies device semantics.
\item Formulate diagnostic coverage as two relations, signal identity and
controller reachability, over a power-aware admissibility model covering series
passives, muxes, level translators, I2C repeaters, and device-mediated
sense-to-I2C monitoring. Union-Find and BFS serve as exact solvers.
\item Emit a proof-path trace card for every verdict, making the method
interpretable by construction and reducing review to the unmonitored set. Because
each card records the connector pin at a board boundary, cards can be joined across
boards for system-level review.
\item Demonstrate, through a cumulative ablation and a cross-model reproducibility
study, that given correct semantics, seeds, and admissibility rules, the
reliability of a monitored verdict comes from the netlist-grounded procedure rather
than any single language model. The verifier guarantees a valid witness path
within the model, and the LLM-supplied semantics determine how many signals are
recovered.
\end{enumerate}
The contribution lies in the DSR-specific formulation of signal identity and admissible observability—not in the underlying graph-search algorithms, which serve as standard computational tools.
The remainder of the paper is organized as follows. Section~\ref{sec:related}
reviews related work and positions NetFlow-DSR against it.
Section~\ref{sec:method} formalizes diagnostic coverage over a netlist graph and
presents the NetFlow-DSR methodology. Section~\ref{sec:results} reports the
experimental results, discusses their implications, and notes the study's
limitations. Section~\ref{sec:conclusion} concludes the paper and outlines
directions for future work.

\section{Related Work}\label{sec:related}
Automated approaches to diagnostic coverage fall into two families, both
insufficient on their own. Net-name heuristics match diagnostic keywords or
controller tokens in signal names. They are fast and training-free, but they
over-claim signals whose names contain misleading tokens and miss signals whose
destination is not encoded in the name. 
  Learning- and clustering-based methods that operate directly
  on netlist or
  board structure, such as gate-level reverse engineering and
  hardware-Trojan
  detection, depend on models fitted to that structure,
  whether trained on
  labeled examples~\cite{alrahis2021gnn,ma2025hardware} or
  fitted without
  supervision~\cite{11342362}, and emit class labels
  rather than an
  auditable controller path, so their residual
  misclassifications cannot be
  verified against the netlist. They
also target different problems, module identification and Trojan detection,
rather than signal-to-controller reachability. Where structure is instead read
from images, node identification is comparatively reliable while edge recovery,
and especially edge direction and polarity, is the persistent weakness, with the
best mean edge-with-polarity F1 at 0.54--0.60 versus 0.99 for nodes in system-map
extraction \cite{white2026using}. Formal verification has been applied at the
netlist level, but to device-level SPICE netlists rather than board-level
observability \cite{jha2025verisim}; broad reviews of LLM-based hardware design
and verification concentrate on RTL and functional flows
\cite{abdollahi2024hardware}; and classical verification relies on simulation and
formal methods \cite{lam2005hardware}, with information-flow coverage formalized
at the register-transfer level \cite{meza2023information}. None of these
addresses board-level diagnostic coverage. That gap is compounded by how
connectivity is often obtained. Many approaches reconstruct it from pixels,
text, or net names, for example by extracting a netlist from circuit images
\cite{peker2026fully,erozan2019reverse}. Schematic-capture tools already emit an
exact electrical graph, so re-deriving it reintroduces perception and inference
errors into a problem whose structural component reduces to explicit algorithmic
steps. Even netlist foundation models that learn
representations directly from the graph \cite{fang2025nettag} produce learned
embeddings rather than an auditable controller path, underscoring the case
for computing connectivity deterministically rather than learning it.

\begin{table*}[!t]
\renewcommand{\arraystretch}{1.15}
\caption{Literature Positioning of NetFlow-DSR Against Related Work}
\label{tab:related}
\centering
\footnotesize
\begin{tabular}{p{2.6cm}p{1.8cm}p{5.4cm}p{5.4cm}}
\toprule
\textbf{Topic / research line} & \textbf{Representative reference(s)} &
\textbf{Key limitation or gap} & \textbf{NetFlow-DSR contribution}\\
\midrule
LLMs for hardware design and EDA (generation, orchestration, comprehension) &
\cite{thakur2024verigen}, \cite{wu2024chateda}, \cite{ye2025llm4netlist} &
The LLM or agent generates HDL or netlists and orchestrates EDA tools;
correctness is judged by simulation, task success, or probabilistic scoring
rather than by a netlist-backed controller proof path, and board-level
signal-to-controller DSR is absent. &
Connectivity is computed from the netlist by deterministic graph algorithms. The
LLM orchestrates the pipeline and supplies device semantics but does not perform
the structural computation, and every positive verdict carries a proof path.\\
\midrule
Vision and image-based structure extraction (image-to-netlist) &
\cite{peker2026fully} &
Structure is reconstructed from pixels or circuit images, so it is perceived
rather than read from an authoritative source. Edge recovery, especially edge
direction and polarity, is the persistent weakness (best mean edge-with-polarity
F1 0.54--0.60 versus 0.99 for nodes), reintroducing perception error into the
connectivity. &
Structural edges are taken directly from the netlist rather than inferred from
images, so no perception step is involved and each monitored verdict traces to
netlist or pin-role evidence.\\
\midrule
Graph learning on netlists (GNN classifiers, netlist representation models) &
\cite{alrahis2021gnn}, \cite{ma2025hardware}, \cite{fang2025nettag} &
Learned models operate on the netlist graph but require training, whether labeled
data or self-supervised pretraining, and target different objectives such as
module identification, Trojan detection, or representation learning. Their
outputs are class labels or learned embeddings whose residual errors carry no
controller proof path. &
Connectivity is computed deterministically over netlist-derived edges rather than
learned or classified, requires no training, and every positive verdict carries
an auditable proof path against the netlist.\\
\midrule
Verification, coverage, and neuro-symbolic checking &
\cite{jha2025verisim}, \cite{abdollahi2024hardware}, \cite{meza2023information} &
Prior work covers RTL and functional verification, as a scoping review confirms,
hardware-security information-flow coverage, and device-level formal verification
of SPICE netlists. None addresses board-level diagnostic coverage or
signal-to-controller reachability with trace-card proof paths. &
Exact graph algorithms compute connectivity over netlist-derived edges and define
board-level diagnostic coverage as controller reachability, keeping the LLM out
of the structural computation while every positive verdict carries an auditable
proof path.\\
\bottomrule
\end{tabular}
\end{table*}

Against this backdrop, NetFlow-DSR occupies a position none of these families
reaches. It keeps the learned and computed parts of the problem separate, using
an LLM only for the device semantics that a netlist cannot encode. Connectivity
is computed by deterministic engines over the netlist's exact edges rather than
inferred from schematics, text, or net names. Unlike net-name
heuristics, it does not depend on destination tokens in signal names; unlike
graph-learning methods, it emits an auditable controller path rather than a class
label; and unlike image-to-netlist pipelines, it does not re-derive a graph the
design tool already provides. Because every positive verdict carries a proof
path, a monitored verdict can be checked directly against the netlist rather than
accepted on the strength of a post hoc rationale, reflecting the
interpretability-by-design developed in explainable AI \cite{arrieta2019explainable}.
Table~\ref{tab:related} positions NetFlow-DSR against representative prior work
spanning LLM-based hardware design, netlist comprehension and generation, graph
learning on netlists, image-to-netlist extraction, and hardware verification. It
highlights the two capabilities that distinguish board-level diagnostic-coverage
verification, namely netlist-grounded structural connectivity and auditable proof
paths.

\section{Methodology}\label{sec:method}
\begin{figure*}[!t]
\centering
\includegraphics[width=1\textwidth]{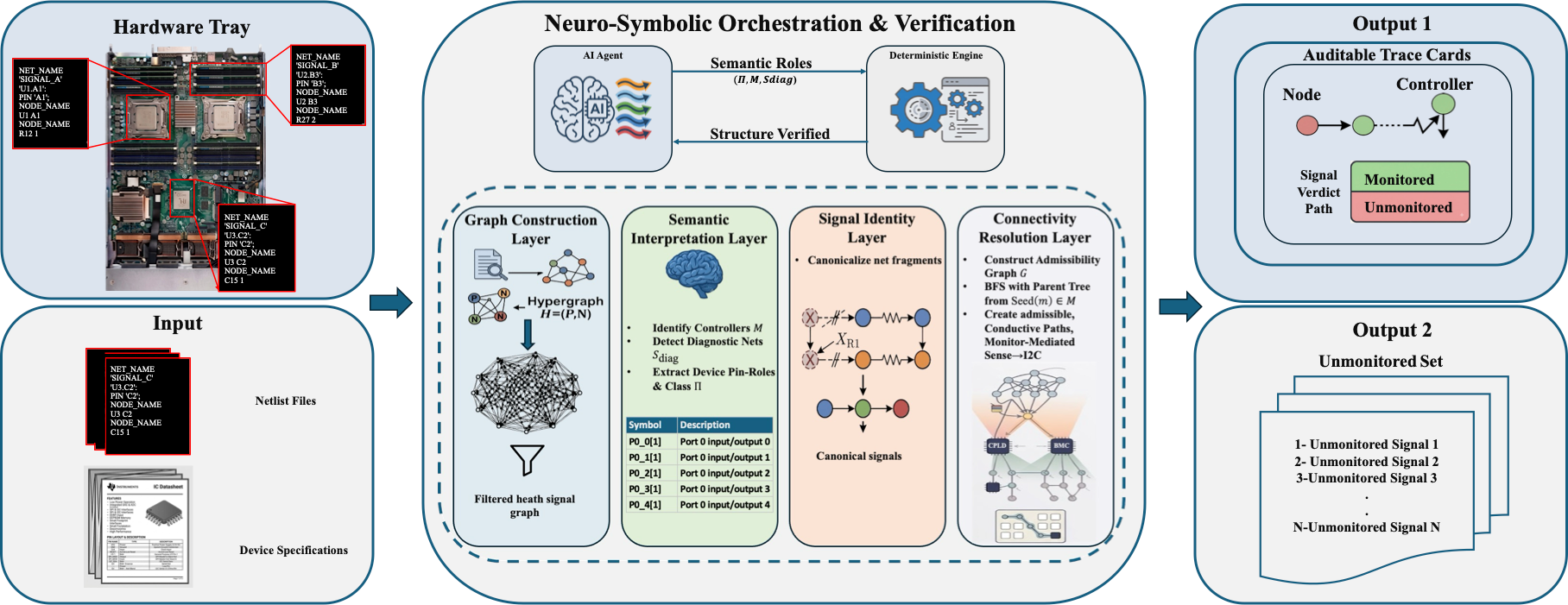}
\caption{NetFlow-DSR verifies Diagnostic Signal Reachability by separating learned semantics from computed structure. An LLM identifies controllers, selects diagnostic signals, and extracts device pin-roles from device specifications; deterministic Union-Find and power-aware BFS then prove connectivity over the authoritative netlist. Each monitored verdict carries an auditable proof-path trace card, yielding a verifier that is sound by construction and interpretable by design.}
\label{fig:abstract}
\end{figure*}

NetFlow-DSR is a netlist-grounded neuro-symbolic pipeline in which an LLM leads
the process end to end, sequencing the stages, invoking deterministic engines as
tools, and supplying their semantic inputs. The connectivity computation and the
assembly of the final audited verdict are carried out by the deterministic graph
engines. The method is organized as four layers. The Graph Construction Layer
parses the raw netlist into a device-centric pinout and builds the power-aware
signal-conduction graph. The Semantic Interpretation Layer identifies management
controllers, in-scope diagnostic signals, and device pin-roles, using the LLM for
the judgments that require device knowledge (controller identification and device
class) and deterministic detectors for the rest (keyword-based signal detection
and library-declared pin directionality). The Signal Identity Layer applies a
Union-Find identity relation that merges series-passive-renamed fragments into
canonical signals. The Connectivity Resolution Layer performs controller-seeded
breadth-first reachability over an admissibility graph that augments identity
with modeled pass-through and device-mediated sense-to-I2C edges, with power rails
excluded. Together, these layers realize a two-relation formulation in which
identity is kept separate from reachability. Doing so avoids two opposite errors:
over-merging distinct multiplexer channels when every conductive path is treated
as identity, and under-tracing legitimate paths when only exact net identity is
considered. The LLM is the orchestrator and semantic front end. It never computes
the structural path itself. Because reachability is decided by breadth-first
search over exact netlist edges and explicit admissible semantic edges, every
monitored verdict is accompanied by an inspectable proof path. Fig.~\ref{fig:abstract}
summarizes the approach.

\subsection{Graph Construction Layer}
The Graph Construction Layer turns the raw netlist into a structured connectivity
model. A board is represented as a netlist hypergraph $H = (P, N)$, where $P$ is
the set of component pins and $N$ is the set of electrical nets. Each net
$n \in N$ is incident to a subset of pins $\mathrm{pins}(n) \subseteq P$, and
$\mathrm{nets}(p)$ denotes the nets incident to pin $p$. Two pins are directly
connected when they share a net. Construction proceeds in two steps, pinout
extraction and graph assembly. Pinout extraction parses the vendor netlist (here,
OrCAD expanded netlists) into device-centric records, one per pin, of the form
(reference designator (refdes), pin, part, net). An example record is
(\texttt{U1}, \texttt{AD15}, \texttt{AST2620}, \texttt{FM\_TPM\_PRSNT\_1\_N}). For
every component, the pinout therefore states which signal each pin connects to. Graph
assembly then inverts this pinout into a net-centric incidence structure that
records, for every net, the pins and components it touches.

Formally, pinout extraction recovers the pin--net incidence $\mathrm{pins}(n)$
and $\mathrm{nets}(p)$ together with a component map
$\mathrm{cmp}(p) \to (\text{refdes}, \text{part})$. The pinout diagram is the
device-centric view keyed by component, and the net-centric incidence is its
equivalent dual. Because extraction applies exact parsing rules, it is lossless
and auditable. Every edge in the resulting graph traces back to a specific pin
record in the netlist, which later supplies the evidence anchors used by the
trace cards.

Nets divide into power nets and signal nets. Power nets distribute a supply
voltage or ground reference, whereas signal nets carry information and are the
subject of DSR. We introduce a structural classifier
\begin{equation}
\mathrm{ispower} : N \to \{0, 1\}
\end{equation}
where $\mathrm{ispower}(n) = 1$ marks net $n$ as a power net (a supply rail or
ground) and $\mathrm{ispower}(n) = 0$ marks it as a signal net. The
$\mathrm{ispower}$ function is a rule-based name classifier applied before
reachability. A net that carries a health or diagnostic token (for example
\texttt{PWRGD}, \texttt{PRSNT}, \texttt{PERST}, \texttt{RESET}, \texttt{ALERT},
or \texttt{SENSE}) is treated as a signal even when its name also resembles a
rail. A net is otherwise marked as power when its name matches conventional
supply or ground forms such as \texttt{GND}, \texttt{VSS}, \texttt{P*V*}, the
\texttt{VDD} family, \texttt{VBAT}, \texttt{VTT}, \texttt{VPP}, \texttt{VIN},
\texttt{VOUT}, and \texttt{VREF}. Excluding these high-fan-out nets prevents
unrelated signals from appearing connected through a shared rail. This predicate
partitions the netlist into power and signal nets,
\begin{equation}
\Npwr = \{ n \in N : \mathrm{ispower}(n) = 1 \}, \quad \Nsig = N \setminus \Npwr
\end{equation}
where $\Npwr$ is the set of power nets, $\Nsig = N \setminus \Npwr$ the signal
nets, and ``$\setminus$'' denotes set difference. This partition is essential for
correct reachability, because power rails and grounds are high-fan-out nets,
allowing traversal through them would make many unrelated signals appear
connected to every controller. NetFlow-DSR therefore excludes power nets as
traversal conductors. A power rail can nonetheless remain a diagnostic endpoint,
credited through a sense-to-I2C or ADC edge rather than by conduction. Power nets
are therefore kept as vertices of the reachability graph $G$ but receive no
identity or pass-through edges; they can only be the endpoint of a sense edge,
so a rail can be reached but never traversed. Signal identity is constructed on
the non-power signal graph,
\begin{equation}
\Gsig = (\Nsig, E), \quad E \subseteq \Nsig \times \Nsig
\end{equation}
where $\Gsig$ is the signal-conduction graph, whose vertices are the signal nets
$\Nsig$, and power nets contribute no edges and therefore act as non-conducting
walls. The Graph Construction Layer fixes this vertex set. The two remaining layers
then populate two distinct edge relations, one at each level of the construction.
The Signal Identity Layer populates $E$ with the series-passive adjacencies
defined in Section~\ref{sec:identity}, so that the connected components of
$\Gsig$ are exactly the canonical signals $S$. The Connectivity Resolution Layer
works one level up, on those components, populating a separate admissible edge set
$\Ereach \subseteq V \times V$, where $V = S \cup \Npwr$, with pass-through and device-mediated edges. Each sense edge is directed from the monitor's I2C signal to the observed net, so the search can enter an observed net from a reachable monitor but never return through that edge. Power nets are sinks, $\deg_G^{+}(r)=0$ for all $r \in \Npwr$, so a rail can be reached but never traversed. The traversal searches $G = (V, \Ereach)$. Each of these operations is
defined explicitly over netlist-grounded inputs, so every edge traces back to a
specific netlist record and remains auditable.

\subsection{Semantic Interpretation Layer}
Operating on the parsed netlist and device inventory produced by the Graph
Construction Layer, the LLM performs a semantic survey of the design and supplies
the inputs consumed by the connectivity engines. It computes no connectivity
itself. It first identifies the management controllers by surveying the component
inventory and flagging candidate BMC, CPLD, FPGA, and system-controller devices
by part function. Known controller families are matched automatically, and the
LLM confirms or resolves ambiguous parts. A direct-node smart-controller
detector, which marks a signal as monitored only when a management controller is
a direct node on its net, constitutes the S1 baseline. In-scope diagnostic
signals are then selected by a deterministic detector that applies the DSR-library
keyword categories to net names, keeping health signals and excluding local
analog feedback nodes, voltage references, compensation nodes, and configuration
straps. The keyword categories are curated with the LLM, but the per-net
selection is a fixed, reproducible rule. Finally, it establishes the device
pin-roles that the connectivity layer consumes. A pin-role has two components, the
electrical directionality of each pin (input, output, bidirectional, or power) and
the functional class of the device (I2C mux, level translator, I2C bus
buffer/repeater, power monitor, ADC, temperature sensor, or GPIO expander).
NetFlow-DSR draws these from two complementary provenances. Directionality is
obtained deterministically from the design's own EDA library, the OrCAD/Allegro
\texttt{LIBRARY\_PARTS} file (\texttt{pstchip.dat}) emitted by the same tool that
writes the netlist, in which each pin declares its role through the
\texttt{INPUT\_LOAD}, \texttt{OUTPUT\_LOAD}, \texttt{BIDIRECTIONAL}, and
\texttt{PINUSE} fields. Because these fields are authored by the schematic-capture
tool, pin direction is read rather than inferred and carries the same authority as
the netlist itself. The functional device class is resolved in order of cost.
Most parts are classified from their part-name family and from the pin names
declared in the same library. For example, an SCL/SDA pair together with channel
pins SC0--SD3 identifies an I2C multiplexer. The device specification (datasheet)
is consulted only for the parts that remain ambiguous, and only for what names and
pins cannot give, such as which analog input senses which rail and whether the
measurement is reported over I2C. A part that no source settles is flagged for its
datasheet rather than guessed. When both sources are
available, the deterministic library takes precedence for directionality and the
device specification supplies only the residual semantics, further narrowing the
part of the pipeline that depends on learned extraction. The two-source device
pin-roles are collected as the pin-role library $\Pi$. Together with the
controller set and the in-scope diagnostic signals, $\Pi$ forms the semantic
input consumed by the downstream deterministic layers.

\begin{figure}[!t]
\centering
\includegraphics[width=0.7\linewidth]{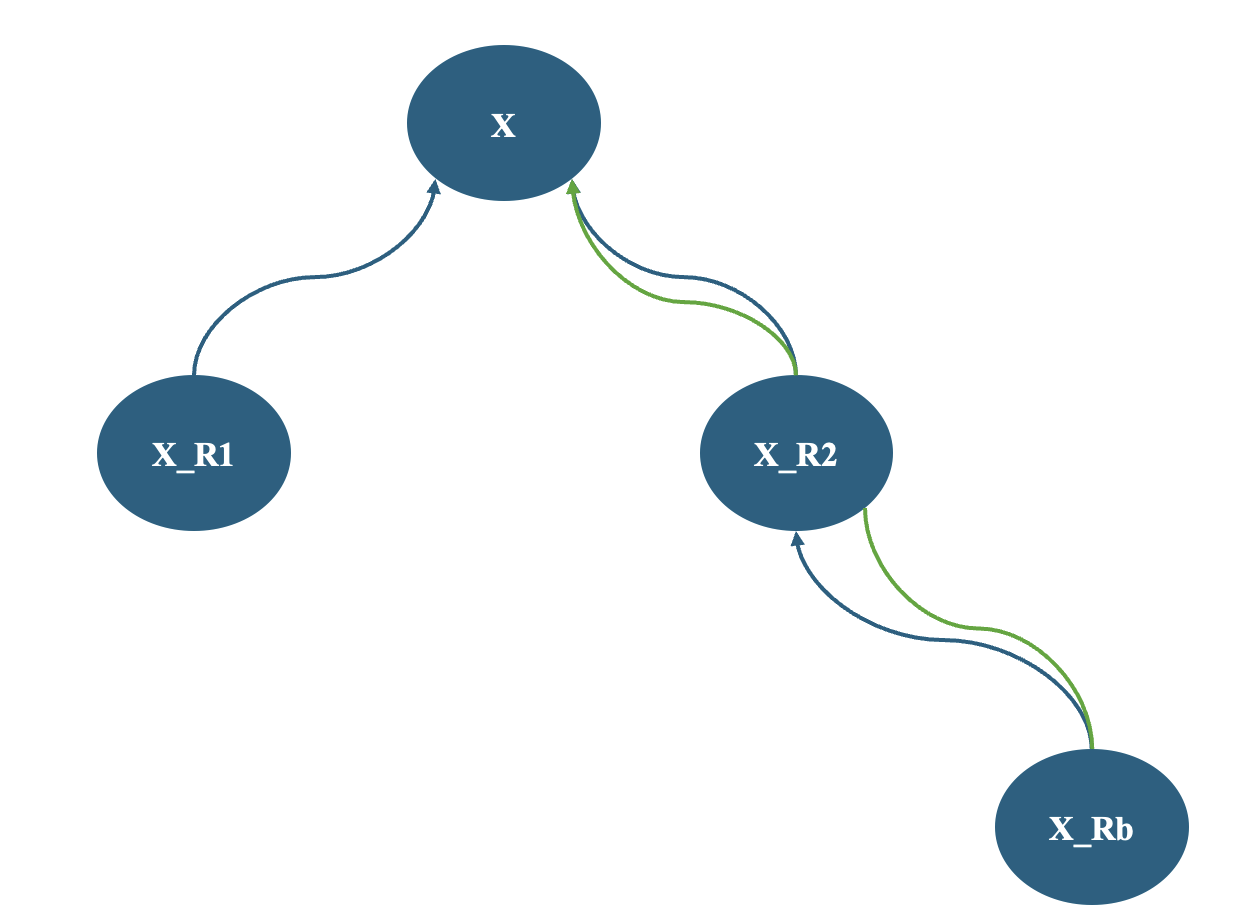}
\caption{Union-Find defragmentation. Blue arrows are the disjoint-set parent
pointers, each net fragment linking to its immediate parent, whereas the green
arrow is the shortcut installed by path compression, re-pointing the deepest
fragment \texttt{X\_Rb} directly to the canonical representative \texttt{X}.}
\label{fig:union}
\end{figure}

\subsection{Signal Identity Layer}\label{sec:identity}
\begin{figure*}[!t]
\centering
\includegraphics[width=\textwidth]{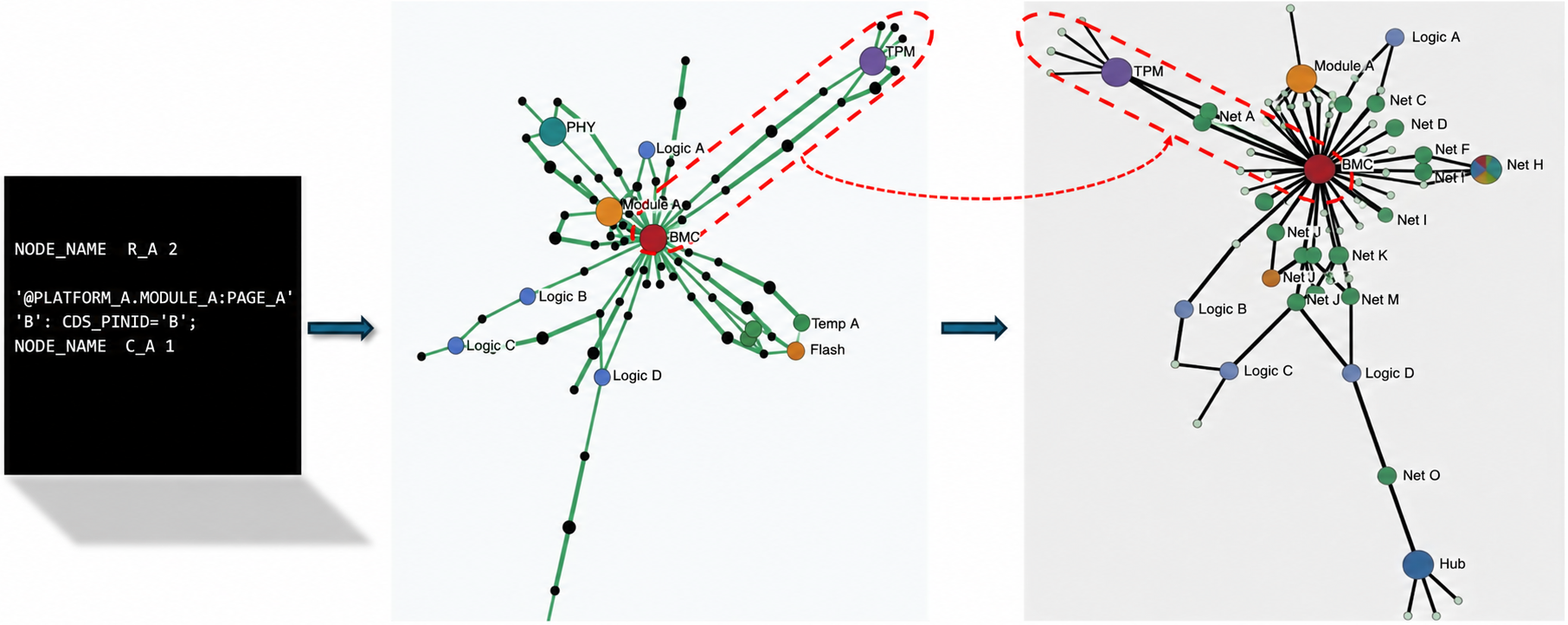}
\caption{\textcolor{red}{Canonicalization of a single configuration, series-passive-renamed net fragments are merged by Union-Find into one canonical signal.}}
\label{fig:canon}
\end{figure*}

The Signal Identity Layer establishes signal identity. A physical signal may
appear under several net names when it crosses a series resistor, ferrite bead, or
zero-ohm link, so a net-by-net analysis, including an LLM trace over schematic
text, can lose the signal at each renaming boundary. This layer resolves the
fragmentation by merging two non-power nets whenever they are joined by an
admissible two-terminal series passive (a passive with exactly two non-power
legs). Chains of such merges define the canonical signal components. As depicted
in Fig.~\ref{fig:union}, canonical components are computed by disjoint-set
Union-Find with union by rank and path compression. A sequence of merges over
the signal nets $\Nsig$ therefore runs in near-linear total time, with amortized
cost $O(\alpha(|\Nsig|))$ per operation, where $\alpha$ is the inverse Ackermann
function. Because $\alpha$ grows so slowly that it stays below five for any input
size arising in practice, canonicalization is effectively constant-time per query
and adds no meaningful cost to the pipeline. We write $s(n)$ for the canonical
representative of net $n$ and $S$ for the set of canonical signals. A pull-up or pull-down to a
power rail is not a merge, because the rail is a power net and is excluded from
the identity relation. Multiplexers, buffers, level translators, and other active
or directional devices are likewise excluded from identity. Their ports may become
reachable under certain conditions, but they are not the same conductor. This
distinction prevents, for example, multiple downstream multiplexer channels from
collapsing into one canonical signal. Fig.~\ref{fig:canon} illustrates the
pipeline of one configuration, from netlist to canonicalization.

\subsection{Connectivity Resolution Layer}
\begin{figure}[!t]
\centering
\includegraphics[width=\linewidth]{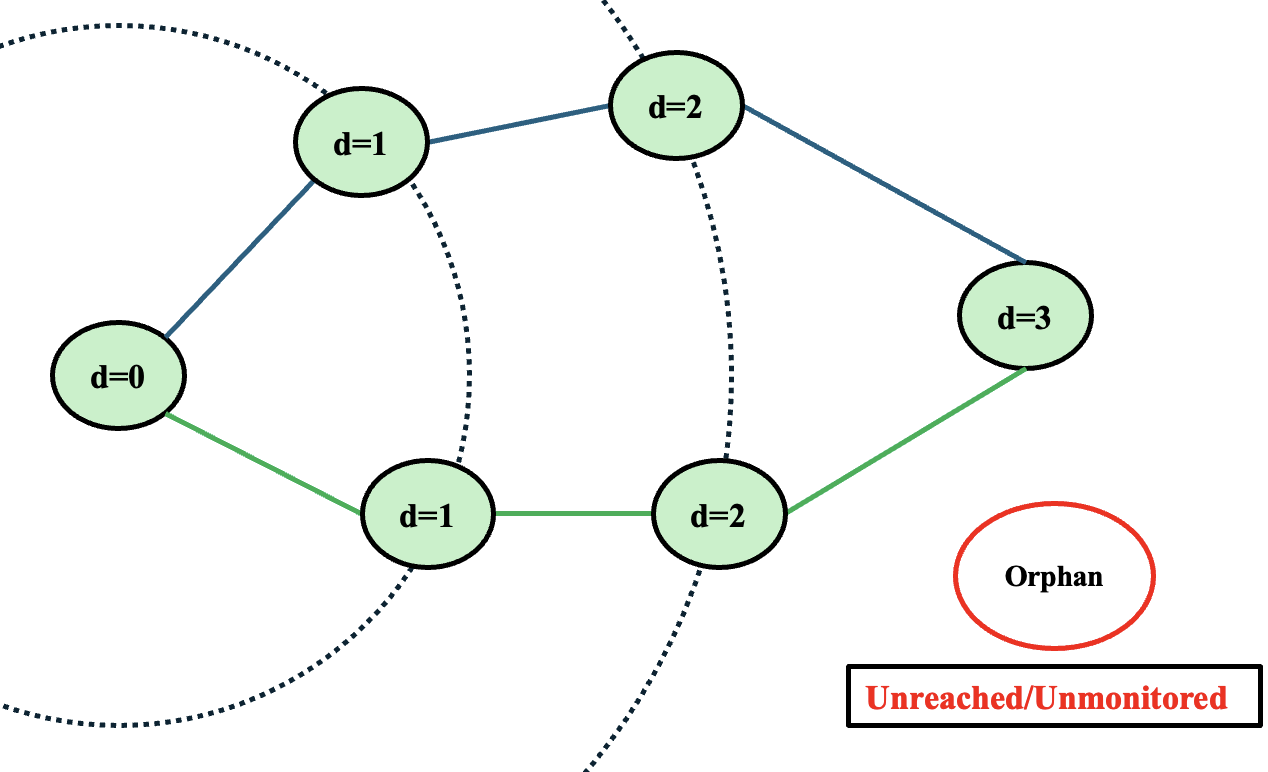}
\caption{BFS model.}
\label{fig:bfs}
\end{figure}

The Connectivity Resolution Layer establishes controller reachability over a graph
larger than identity. Coverage is evaluated relative to a set of management
controllers $M$, such as on-board CPLDs or BMCs, or connector egresses that lead
to off-board management devices; each controller $m \in M$ contributes a set of
seed pins $\mathrm{seed}(m) \subseteq P$. Because the reachability graph is defined
over canonical signals rather than pins, these pins are first lifted to the set of
seed signals
$\sigma(m) = \{ s(n) : p \in \mathrm{seed}(m),\, p \in \mathrm{pins}(n),\, \mathrm{ispower}(n) = 0 \}$,
which is the set actually used to seed the traversal.
Because identity has already been absorbed into the vertices of $G$, two
canonical signals are adjacent in the reachability graph only when they are joined
by a modeled pass-through device, namely an I2C multiplexer channel, an I2C bus
buffer/repeater pass-through, or a level-translator channel, or by a
device-mediated sense-to-I2C edge.
Power rails are excluded from traversal, preventing pull-up networks from creating
false controller paths. Pin directionality, taken from the design library, further
constrains admissibility. Each pass-through or device-mediated edge is added
consistently with the input, output, or bidirectional role of the pins it joins.
\textcolor{black}{For a purely combinational device whose
output is reachable from a controller, a backward relation
to an input represents a potential influence on the monitored
output, not necessarily independent observability of that
input. Pin directionality establishes the direction of this
relation but is insufficient to establish observability.
Crediting an input as monitored additionally requires
device-function and operating-condition evidence that the
input's diagnostic state is observable at the output.} The traversal
therefore runs BFS over $G$, whose vertices carry the identity relation and whose
edges carry the admissible device-hop relation. Fig.~\ref{fig:bfs} illustrates
this composed traversal.

  \begin{equation}
  T_m = \mathrm{BFS}(G, \sigma(m)), \quad \mathrm{cov}(s, m) =
  \mathds{1}[\,s \in T_m\,]
  \end{equation}
  
where $G = (V, \Ereach)$ is the reachability graph, over canonical signals and power-net endpoints,,
$\mathrm{BFS}(G, \sigma(m))$ is breadth-first search on $G$ from the seed signals
$\sigma(m)$, $T_m \subseteq V$ is the resulting reachable set, and
$\mathds{1}[\cdot]$ is the indicator function. Thus $\mathrm{cov}(s, m) = 1$ exactly when signal $s$ is
monitored, that is, reachable, by controller $m$. Through device-mediated edges,
coverage also credits signals observed by a monitor device (e.g., a power monitor
or ADC) that has no direct electrical path to the controller. Reachable sets are
maintained separately for each controller, so a signal can be classified as
BMC-only, CPLD-only, both, or neither. The diagnostic signal set $\Sdiag \subseteq V$ excludes local analog feedback and
reference nodes such as \texttt{*\_FB}, \texttt{*\_REF}, and \texttt{*\_SW}
local-regulator nodes. For each controller, the DSR output is the monitored set
and its complement, the unmonitored set,
\begin{equation}
\mathrm{Mon}_m = \{s \in S_{\mathrm{diag}} : \mathrm{cov}(s, m) = 1 \}
\end{equation}
\begin{equation}
A_m = S_{\mathrm{diag}} \setminus \mathrm{Mon}_m
\end{equation}
where $\mathrm{Mon}_m$ is the set of diagnostic signals monitored by controller
$m$ and $A_m$ is the corresponding unmonitored set. The unmonitored set $A_m$ is
the primary review output, it enumerates the diagnostic signals for which no
admissible management path has been proven. For each controller $m$, NetFlow-DSR
performs a breadth-first flood from $\sigma(m)$ and records the reachable
set. Each controller-seeded flood runs in $O(|S| + |\Ereach|)$ time and produces
both a verdict and a parent tree, which is used as the proof path for every
reached signal. The algorithmic novelty is not BFS itself, but the DSR-specific
admissibility predicate that determines which signal-to-signal hops are valid,
which rails are barriers, and which functional device observations can be credited
as monitored. The soundness of this construction can be stated precisely.

\begin{proposition}[Witness soundness]
\label{prop:sound}
If $\mathrm{cov}(s,m) = 1$, then the parent tree returned by
$\mathrm{BFS}(G, \sigma(m))$ exhibits a path in $G$ from a seed signal in
$\sigma(m)$ to $s$ every edge of which is admissible.
\end{proposition}

  \begin{figure}[!t]
  \centering
  \includegraphics[width=1\linewidth]{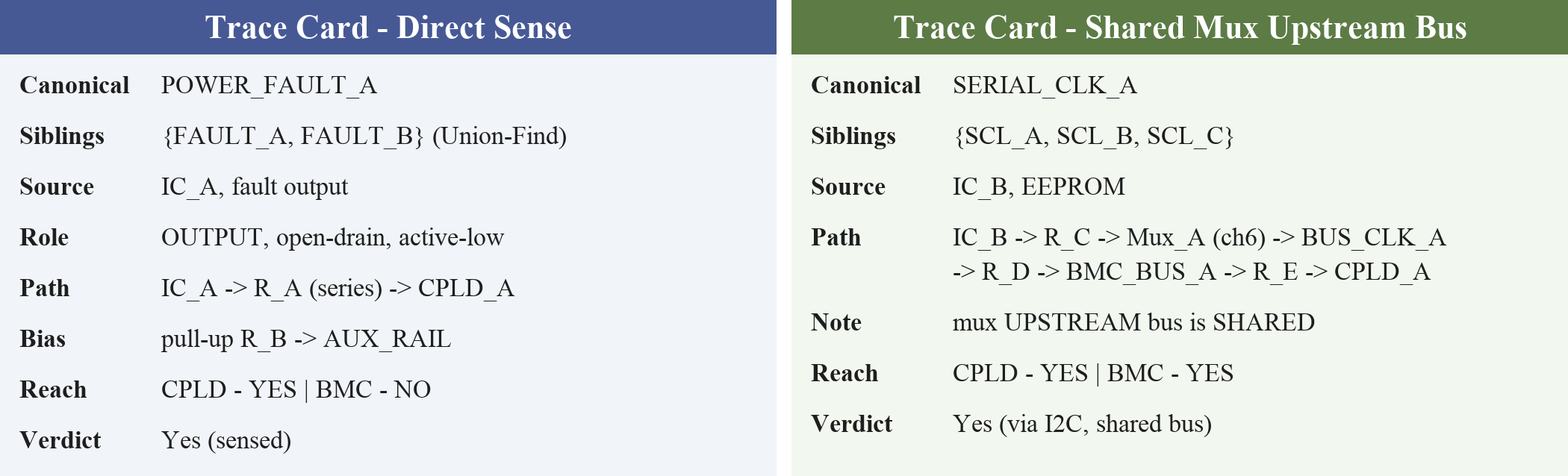}
  \caption{\textcolor{red}{Per-signal trace cards, the LLM-grounding unit.}}
  \label{fig:cards}
  \end{figure}
  
\noindent The proof is immediate from the parent tree, which records for each
reached vertex the admissible edge by which it was reached. Two consequences
follow, and it is worth separating them. Proposition~\ref{prop:sound} is internal
to the admissibility model, and in that sense the verifier never reports a signal
as monitored without an explicit witness path. External validity, namely that
every admissible edge corresponds to real observability, is an empirical property
of the edge set rather than a theorem. A false positive can therefore arise only
from an incorrect admissible edge, for example a mis-supplied device class, and
never from the search itself. Soundness in this sense does not eliminate the
trust question, it relocates it. Instead of asking whether a verdict is correct,
a reviewer asks whether one edge is admissible, a local question that the trace
card answers directly. The 100\% precision observed in the evaluated data is
consistent with this structure. The engines are deterministic and, with the hash seed
fixed, bit-reproducible, so identical inputs yield identical verdicts and proof
paths. Device-mediated monitoring extends the admissibility graph for monitor
devices. Some rails or pins do not connect electrically to a controller but are
observed by a device such as a power monitor, ADC, or GPIO expander, which relays
the measurement or pin state over I2C. In this case, the observed net is credited
as indirectly monitored only if the device observes that net and the device I2C
bus itself reaches the controller. A GPIO expander's input pins are read in
exactly this way, through the expander's I2C port register. Temperature and leak
sensors that report internal measurements require no external sense edge. Their
coverage is decided by reachability of their I2C and alert signals. This rule adds
semantic edges derived from device function without changing the underlying graph
search. A trace card is the unit of evidence produced for each canonical signal.
It records the merged net fragments, the source leaf and associated part, the
traced path to each controller, the role of each modeled hop, and the resulting
verdict. Fig.~\ref{fig:cards} shows representative cards. In one, a fault signal
reaches the CPLD through a series resistor, while an SMBus clock reaches both CPLD
and BMC through a shared mux path. These cards make positive verdicts auditable and expose the unmonitored set for review.
Trace cards also carry signal identity beyond a single board. When a controller is
declared as a connector egress, the card records the exact connector pin at which the
signal leaves the board, and the card of the mating board starts at the same pin. Cards
from two boards can therefore be joined at that pin, so a signal is followed across
boards by matching its cards rather than by re-tracing it on each board. Algorithm~\ref{alg:netflow} summarizes
the complete NetFlow-DSR procedure over the notation of Section~\ref{sec:method}.

\begin{algorithm}[!t]
\caption{NetFlow-DSR: Netlist-Grounded Diagnostic-Coverage Verification}
\label{alg:netflow}
\begin{algorithmic}[1]
\Require netlist hypergraph $H=(P,N)$; pin-role library $\Pi$; controller set $M$ with seed pins $\mathrm{seed}(m) \subseteq P$
\Ensure coverage $\mathrm{cov}(\cdot,\cdot)$; unmonitored sets $\{A_m\}$; per-signal trace cards $C$
\Function{NetFlow-DSR}{$H,\Pi,M$}
  \State $\Nsig \gets \{n \in N : \mathrm{ispower}(n)=0 \}$ \Comment{power nets are not conductors}
  \State initialize disjoint-set forest over $\Nsig$ \Comment{Signal Identity Layer}
  \For{each two-terminal series passive with legs $(a,b)$, both non-power}
     \State \textsc{Union}$(a,b)$ \Comment{populates $E$ of $\Gsig$}
  \EndFor
  \State $S \gets \Nsig / \textsc{Find}(\cdot)$ \Comment{canonical signals: components of $\Gsig$}
 \State $V \gets S \cup \Npwr$;\quad $\Sdiag \gets \textsc{Detect-Diagnostic}(V)$ \Comment{DSR category keywords}
  \State $\Ereach \gets$ pass-through, I2C$\to$observed-net and combinational back-edges licensed by $\Pi$, oriented by the declared pin directionality \Comment{Connectivity Resolution Layer}
  \State $G \gets (V,\Ereach)$ \Comment{quotient of $\Gsig$; identity is in the vertices}
  \For{each controller $m \in M$}
     \State $\sigma(m) \gets \{ s(n) : p \in \mathrm{seed}(m),\, p \in \mathrm{pins}(n),\, \mathrm{ispower}(n)=0 \}$
     \State $T_m \gets \textsc{BFS}(G,\sigma(m))$ with parent tree \Comment{parent tree $=$ proof path}
     \State $\mathrm{cov}(s,m) \gets \mathds{1}[s \in T_m]$ for all $s \in V$
     \State $\mathrm{Mon}_m \gets \{ s \in \Sdiag : \mathrm{cov}(s,m)=1 \}$;\quad $A_m \gets \Sdiag \setminus \mathrm{Mon}_m$
  \EndFor
  \State $C \gets \{\textsc{Trace-Card}(s) : s \in V\}$ \Comment{Grounding}
  \State \Return $\mathrm{cov}, \{A_m\}, C$
\EndFunction
\end{algorithmic}
\end{algorithm}

\section{Results}\label{sec:results}
NetFlow-DSR is evaluated on an expert-labeled benchmark spanning the principal functional classes that compose a modern AI accelerator server tray, namely host compute,
acceleration, storage, I/O expansion, and tray-level management. This setup
assesses diagnostic-coverage behavior across distinct board roles rather than on a
single board type. The evaluation uses five anonymized production server and
accelerator boards with distinct management-controller architectures. The MB has
an off-board BMC path through an I2C multiplexer and an on-board CPLD. The
Accelerator has an on-board BMC and CPLD. The remaining three boards are a HDD, a
PCIe board, and a SCM. Together, these boards cover direct controller wiring,
renamed nets across series passives, pass-through devices, and device-mediated
telemetry. Annotation was performed at the signal level. Candidate diagnostic
signals were collected from netlist names and DSR categories, including
power-good, presence, reset, interrupt, alert, I2C/SMBus, ADC, fault, enable, and
telemetry signals. Local analog feedback nodes, voltage references,
regulator-compensation nodes, configuration straps, and duplicate aliases judged
by manual review to refer to the same physical signal were excluded from scope.
Ground truth was assigned by three hardware engineers experienced in server
platform design review. For each board, one engineer labeled every in-scope
signal from the schematic, netlist, and the pinout descriptions of the involved
ICs, and a second engineer verified every label against the same sources, with
differences resolved by joint schematic review. Each in-scope signal received a
binary expert label. A positive label means that manual engineering review
judged the signal observable or controllable by an in-scope management
controller. That path could be a direct electrical one, an admissible
pass-through path, or
device-mediated monitoring supported by the IC pinout descriptions, such as
sense-to-I2C reporting. A negative label means that no such management path was
present within the declared controller scope. These labels were assigned
independently of the NetFlow-DSR output (from the schematic, netlist, and IC
pinout descriptions rather than from the method's trace cards) and were frozen
before any stage was scored.

\begin{figure*}[!t]
\centering
\includegraphics[width=0.9\textwidth]{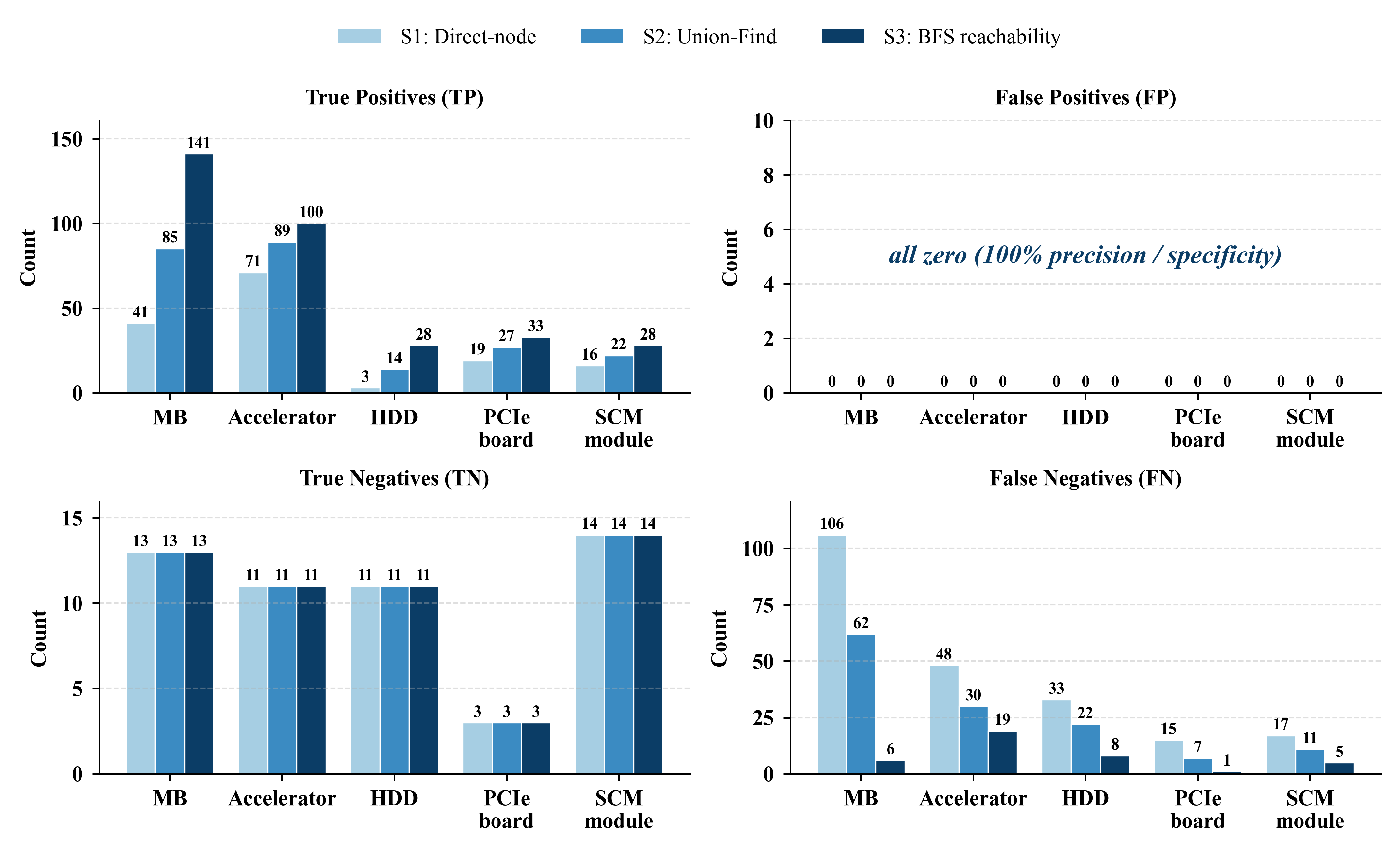}
\caption{Per-board confusion-matrix components (TP, FP, TN, FN) across the three ablation stages.}
\label{fig:confusion}
\end{figure*}

The method is evaluated as a cumulative three-stage ablation. S1 is a direct-node
smart-controller detector that flags a signal as monitored only when a management
controller is a direct node on its net. S2 adds the Signal Identity Layer to apply
canonicalization, and merges renamed net fragments across series passives into
canonical signals, so a controller that is a direct node on any fragment is
credited with monitoring the whole signal. S3 adds the Connectivity Resolution
Layer, which floods the canonical-signal graph from each controller and reaches
signals through pass-through devices (muxes, level translators, I2C repeaters) and
device-mediated sense-to-I2C monitors. The stages are strictly nested
($\mathrm{S1} \subseteq \mathrm{S2} \subseteq \mathrm{S3}$), so each stage can
only add monitored verdicts, never retract them. After in-scope filtering, the
benchmark contains 421 expert-graded signals, of which 369 are positives
(monitored) and 52 are negatives (unmonitored) at the signal level, not the
net-fragment level. This signal count reflects the review scope adopted here, not a limit of the
method. The identity and reachability engines process every net on the board, and
a trace card is produced for every canonical signal, so a team with a different
target, or a designer who needs every signal, obtains verdicts for the full set by
changing only the scope filter. The same fixed labels are used for every stage, so the
differences across stages reflect only the evidence and graph relations available
to the method, not changes in the annotation set. Fig.~\ref{fig:confusion} shows
the resulting confusion-matrix components (TP, FP, TN, FN) per board and stage.
The false-positive count is zero for every board at every stage, while true
positives rise and false negatives fall. Because no false positives occur in the
evaluated data, observed precision and specificity are 100\%. With zero false
positives among 52 negatives, the rule of three bounds the false-positive rate
above by 5.8\% at 95\% confidence, corresponding to specificity of at least 94.2\%
under this sample size, assuming approximately independent negatives from a single platform. The discriminative variation across stages is therefore
carried by accuracy, sensitivity (recall), F1, and MCC, defined in
(\ref{eq:acc})--(\ref{eq:mcc}).

\begin{equation}\label{eq:acc}
\mathrm{Accuracy} = \frac{TP+TN}{TP+TN+FP+FN}
\end{equation}
\begin{equation}\label{eq:prec}
\mathrm{Precision} = \frac{TP}{TP+FP}
\end{equation}
\begin{equation}\label{eq:sens}
\mathrm{Sensitivity\ (Recall)} = \frac{TP}{TP+FN}
\end{equation}
\begin{equation}\label{eq:spec}
\mathrm{Specificity} = \frac{TN}{TN+FP}
\end{equation}
\begin{equation}\label{eq:f1}
F_1 = \frac{2\,TP}{2\,TP+FP+FN}
\end{equation}
\begin{equation}\label{eq:mcc}
\mathrm{MCC} = \frac{TP\cdot TN - FP\cdot FN}{\sqrt{(TP+FP)(TP+FN)(TN+FP)(TN+FN)}}
\end{equation}

\begin{table}[!t]
\renewcommand{\arraystretch}{1.2}
\caption{Per-Signal Verdict Accuracy of the Three-Stage Ablation on Five Production Boards}
\label{tab:persignal}
\centering
\footnotesize
 \begin{tabular}{>{\raggedright\arraybackslash}p{1.38cm}>{\centering\arraybackslash}p{0.7cm}>{\centering\arraybackslash}p{0.7cm}>{\centering\arraybackslash}p{0.7cm}>{\centering\arraybackslash}p{0.7cm}>{\centering\arraybackslash}p{0.7cm}>{\centering\arraybackslash}p{0.7cm}}
  
\toprule
 \textbf{Board} & \textbf{Stage} & \textbf{Acc.\ (\%)} & \textbf{Sens.\ (\%)}
  & \textbf{Prec.\ (\%)} & \textbf{F1} & \textbf{MCC}\\
  \midrule
  \multirow{3}{*}{Motherboard} & S1 & 33.8  & 27.9  & 100 & 0.436 & 0.175\\
                               & S2 & 61.25 & 57.82 & 100 & 0.733 & 0.317\\
                               & \textbf{S3} & \textbf{96.3} & \textbf{95.9} &
  \textbf{100} & \textbf{0.979} & \textbf{0.810}\\
  \midrule
  \multirow{3}{*}{Accelerator} & S1 & 63.1  & 59.7  & 100 & 0.750 & 0.334\\
                               & S2 & 76.9  & 74.8  & 100 & 0.856 & 0.448\\
                               & \textbf{S3} & \textbf{85.4} & \textbf{84.0} &
  \textbf{100} & \textbf{0.913} & \textbf{0.555}\\
  \midrule
  \multirow{3}{*}{HDD}         & S1 & 29.8  & 8.3   & 100 & 0.154 & 0.144\\
                               & S2 & 53.2  & 38.9  & 100 & 0.560 & 0.360\\
                               & \textbf{S3} & \textbf{83.0} & \textbf{77.8} &
  \textbf{100} & \textbf{0.875} & \textbf{0.671}\\
  \midrule
  \multirow{3}{*}{PCIe board}  & S1 & 59.5  & 55.9  & 100 & 0.717 & 0.305\\
                               & S2 & 81.1  & 79.4  & 100 & 0.885 & 0.488\\
                               & \textbf{S3} & \textbf{97.3} & \textbf{97.1} &
  \textbf{100} & \textbf{0.985} & \textbf{0.853}\\
  \midrule
  \multirow{3}{*}{SCM module}  & S1 & 63.8  & 48.5  & 100 & 0.653 & 0.468\\
                               & S2 & 76.6  & 66.7  & 100 & 0.800 & 0.611\\
                               & \textbf{S3} & \textbf{89.4} & \textbf{84.8} &
  \textbf{100} & \textbf{0.918} & \textbf{0.791}\\
  \bottomrule
  \end{tabular}
  \end{table}

Table~\ref{tab:persignal} reports the derived per-signal metrics for each board
(MB $N=160$, Accelerator $N=130$, HDD $N=47$, PCIe board $N=37$, SCM module
$N=47$) and each of the three cumulative stages. Fig.~\ref{fig:confusion} reports
the underlying confusion counts. The S1-to-S2 step produces a large recall gain,
recovering 87 signals ($+23.6$ points of pooled recall), which demonstrates the
importance of canonicalization. Raw netlists often split one physical signal into
multiple named fragments at series resistors, ferrites, or zero-ohm links, and a
direct-node detector misses a signal whenever the controller-side fragment carries
a different name from the reviewed fragment. Union-Find canonicalization merges
these fragments before reachability is evaluated. On the Motherboard, for example,
6{,}196 raw nets (5{,}903 of them signal nets, after excluding 293 power rails)
are reduced to 4{,}803 canonical signals through 1{,}100 fragment merges,
including 710 multi-fragment signals. Recall increases on every board after this
step, from 27.9\% to 57.8\% on the Motherboard, 59.7\% to 74.8\% on the
Accelerator, 8.3\% to 38.9\% on the HDD, 55.9\% to 79.4\% on the PCIe board, and
48.5\% to 66.7\% on the SCM module. The S2-to-S3 improvement adds
controller-seeded reachability, recovering a further 93 signals ($+25.2$ points of
pooled recall). S2 credits a signal only when a controller is a direct node on its canonical
signal. S3 instead runs breadth-first search from each controller over the
canonical-signal graph, flooding across modeled pass-through devices (muxes,
level translators, I2C repeaters) and device-mediated sense-to-I2C relations.
Those relations derive from the device pin-roles the LLM extracted from the
components' device specifications. Reachability at S3 is therefore BFS over the union of the
canonical-signal identity edges from S2 and these admissible pass-through and
device-mediated edges. This effect is most visible on the HDD, where recall
increases from 38.9\% to 77.8\% and F1 from 0.560 to 0.875, and on the PCIe board,
where S3 reaches 97.1\% recall and 0.985 F1. The majority of the remaining false
negatives are power rails. NetFlow-DSR credits a rail as monitored only when a
monitor or ADC directly senses it and relays the measurement over I2C. Power rails
that are monitored through indirect mechanisms the method does not model, for
example a power-good indication from their voltage converter, are not captured by
the current approach and therefore remain false negatives. The remaining false
negatives involve active buffers, which are single-direction and therefore not
bridged, I2C I/O-expander cases not resolved in the evaluated run, or signals whose expert-confirmed endpoint was not declared as a controller seed in the run.

The zero-false-positive result is what makes this recovered recall useful in
practice, because in a diagnostic-coverage sign-off the two error types are not
symmetric. A false negative returns an already-monitored signal to the review
list, adding bounded, self-correcting effort. A false positive, by contrast,
removes a genuinely unmonitored signal from the set of candidate coverage gaps and
lets a fault path escape review silently. Since the negative class is small, only
52 of the 421 signals, this unmonitored set receives the reviewer's full
attention. Even a single over-claim would be disproportionately costly.
NetFlow-DSR is deliberately biased against over-claiming for this reason. An
admissible edge is added only when it is backed by netlist evidence or by a device
pin-role the LLM extracted from a device specification. The 52 negatives therefore
serve as a stress test of soundness in the evaluated data, and the method leaves
all of them correctly unmonitored across every board.
Review effort is redirected from exhaustive signal tracing to focused examination
of the unmonitored set, and each monitored verdict can be audited directly through
its trace-card proof path. On the Accelerator, 100 of 130 signals are returned as
monitored and 30 remain for audit. On the PCIe board, only 4 of 37 remain. This
auditability is a property of the structure itself. The procedure decomposes DSR
into explicit, netlist-grounded stages, and every positive verdict carries a proof
path that can be checked against the netlist. Because the evaluated signal set is
class-imbalanced, MCC (reported in Table~\ref{tab:persignal}) is included as a
balance-aware summary metric alongside accuracy and F1. The dominant error mode of
the direct-node baseline is under-claiming rather than over-claiming. Signals are
missed when their evidence is split across renamed fragments, series passives,
pass-through devices, or monitor-mediated telemetry paths. NetFlow-DSR recovers
these signals by extending the graph only with admissible, evidence-backed edges
over a power-guarded graph, so each stage adds true positives while preserving
empirical zero false positives in the evaluated data.

\begin{figure}[!t]
\centering
\includegraphics[width=0.758\linewidth]{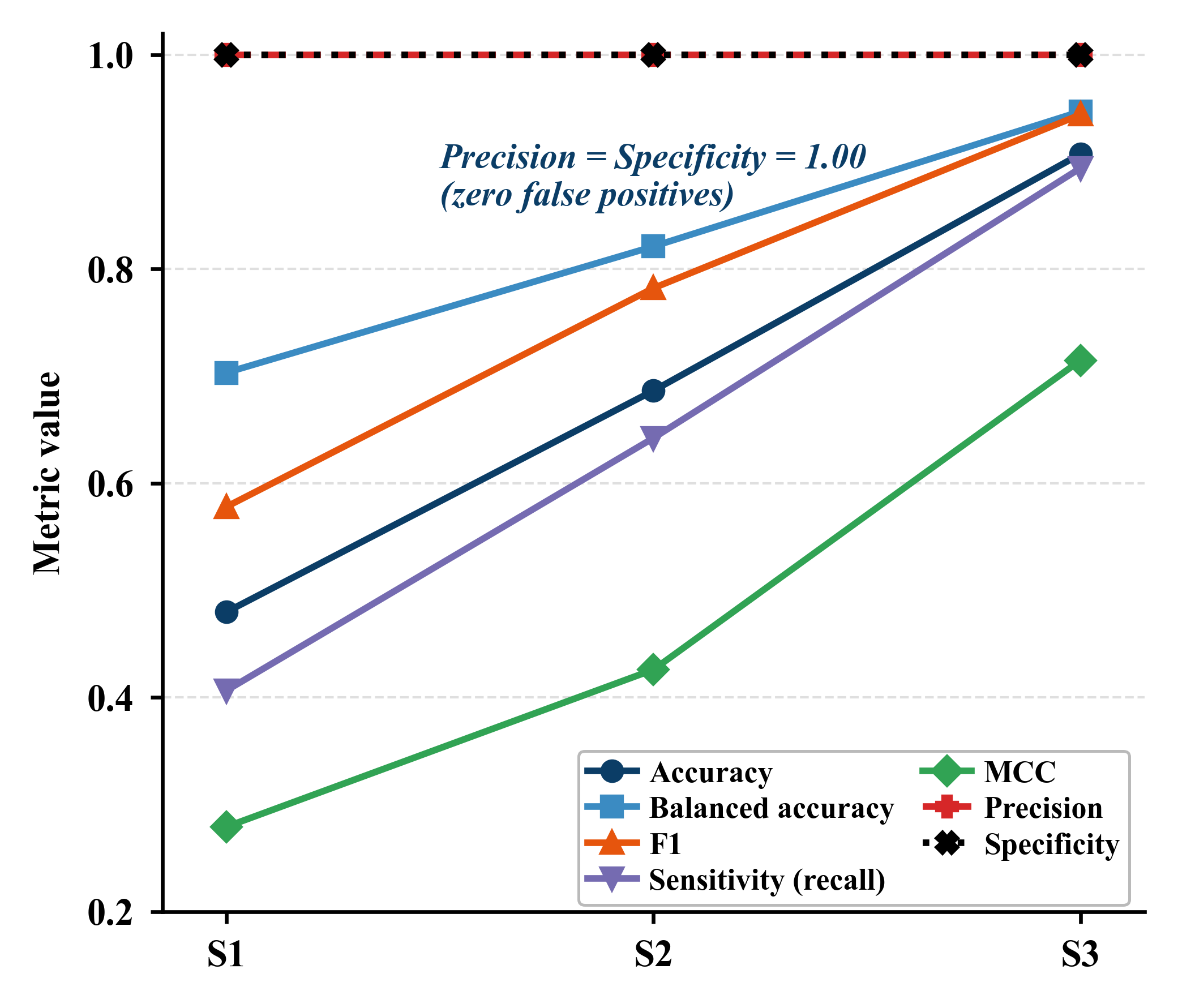}
\caption{Pooled verdict metrics across the three ablation stages.}
\label{fig:pooled}
\end{figure}

Fig.~\ref{fig:pooled} summarizes the pooled tray-level metrics from S1 to S3. Two
regularities hold across every board. First, recall increases monotonically from
S1 through S3, confirming that each added relation, canonicalization at S2 and
device-mediated resolution at S3, recovers monitored signals that the preceding
stage had missed. Second, observed precision and specificity remain fixed at 100\%
at every stage and on every board, so no stage introduces a spurious monitored
verdict in the evaluated data. Pooled across all 421 signals, recall rises from
40.7\% at S1 to 64.2\% at S2 and 89.4\% at S3, while F1 improves from 0.578 to
0.782 and 0.944 and MCC from 0.279 to 0.426 and 0.715. Because observed precision
is held at unity, the F1 gain is attributable entirely to recovered true
positives. The cumulative stages enlarge the set of correctly identified monitored
signals without weakening the reliability of a monitored verdict in this
benchmark. Balanced accuracy, the mean of sensitivity and specificity shown in
Fig.~\ref{fig:pooled}, rises with recall from 0.70 at S1 to 0.82 at S2 and 0.95 at
S3, because specificity is fixed at 100\%. These results give empirical weight to
the perception bottleneck reviewed in Section~\ref{sec:related}. In NetFlow-DSR,
perception-based edge recovery is not used for structural connectivity because the
netlist encodes the electrical edges, connectivity is computed over
netlist-derived structure rather than perceived from pixels. The missed-edge
failures that dominate image-based extraction are therefore removed from the
structural part of the DSR verdict. For the coverage decision used here,
reachability is modeled over admissible monitoring paths. Polarity and signal
direction are semantic properties recorded separately from the binary question of
whether a controller can observe the signal. NetFlow-DSR thus removes the
connectivity portion of the perception bottleneck, which helps explain why
observed precision does not degrade with board complexity or across models. A
further finding is that interpretability is produced by the same mechanism that
delivers accuracy. Because connectivity is computed on the netlist, every
positive verdict is emitted with an explicit proof path, namely the chain of
canonical signals and admissible edges terminating at the controller. Each
verdict is therefore auditable by construction rather than rationalized after
the fact. This realizes the interpretability-by-design reviewed in Section II,
and it is decisive at production scale, where a single board carries several
hundred candidate diagnostic signals and a free-form model rationale would be
infeasible to audit. The remaining disagreements point to extensions rather than
to a change in formulation. Active buffers and scope-dependent endpoints such as
CPU-hosted DIMM telemetry require additional device semantics or broader
controller declarations. The procedure verifies reachability. Whether an
unmonitored signal is an actual design gap remains a semantic engineering
judgment.

   This asymmetry admits a stronger statement than the error counts alone convey. Because the verifier emits no monitored verdict without a netlist-verified proof path, the absence of false positives in the evaluated data implies $\Gamma \subseteq A_m$, where $\Gamma$ denotes the set
  of genuinely unmonitored diagnostic signals and $A_m$ the flagged
  unmonitored set of controller $m$. Every genuinely unmonitored signal in the
  evaluated data is contained in the set the method returns for review.
  Examination of $A_m$ is therefore exhaustive with respect to the genuinely
  unmonitored set rather than a sample of it, and the residual error mode is
  redundant effort rather than escaped risk. This containment property is
  what separates a sound verifier from a merely accurate classifier. A method
  with comparable $F_1$ but a nonzero false-positive rate would discard an
  unknown subset of $\Gamma$ before a reviewer ever saw it, and no aggregate
  metric would reveal which signals had been lost.

  Containment also makes the review burden quantifiable. The quantities that
  follow are reported over the pooled benchmark rather than for one controller,
  so a pooled review set is required. A signal needs review only when no in-scope
  controller reaches it, which is the intersection rather than the union of the
  per-controller sets,
  \begin{equation}
  A \;=\; \bigcap_{m \in M} A_m \;=\; S_{\mathrm{diag}} \setminus
  \bigcup_{m \in M} \mathrm{Mon}_m
  \label{eq:pooled}
  \end{equation}
  where $A$ collects the diagnostic signals left unmonitored by every controller
  on every board. Since $\Gamma \subseteq A_m$ holds for each $m$, containment
  carries over to the pooled set, $\Gamma \subseteq A$. Let the review
  reduction be
  \begin{equation}
  \rho = 1 - \frac{|A|}{|S_{\mathrm{diag}}|}
  \label{eq:reduction}
  \end{equation}
  and let the enrichment $\eta$ be the ratio of unmonitored-signal density within
  $A$ to that density within the full in-scope set $S_{\mathrm{diag}}$.
  These two quantities are not independent under containment, because
  $\Gamma \subseteq A$ fixes the numerator of both densities at $|\Gamma|$, giving
  \begin{equation}
  \eta = \frac{|\Gamma| \,/\, |A|}{|\Gamma| \,/\, |S_{\mathrm{diag}}|}
       = \frac{|S_{\mathrm{diag}}|}{|A|}
       = \frac{1}{1-\rho}
  \label{eq:enrichment}
  \end{equation}
  so the concentration of genuinely unmonitored signals improves at precisely
  the rate at which the review set contracts. Whether a genuinely unmonitored
  signal constitutes a design gap is a separate engineering judgment, so $\rho$
  and $\eta$ quantify the concentration of the review set rather than defect
  density. A verifier that admitted false positives
  would break this identity, because each over-claim removes a member of
  $\Gamma$ from $A$ and reduces the numerator that~\eqref{eq:enrichment}
  holds fixed. Enrichment and effort reduction coincide only for a sound
  procedure.

  Applied to the pooled benchmark, S3 returns 330 of the 421 in-scope signals as
  monitored and flags $|A| = 91$ for review, comprising the 52 true negatives
  and the 39 signals that S3 fails to credit, giving $\rho = 78.4\%$ and
  $\eta = 4.63$. An engineer inspecting the flagged set encounters a genuinely
  unmonitored signal in $57.1\%$ of cases, against a base rate of $12.4\%$ over
  the full signal set, and all 52 remain within it. The corresponding values
  are $\rho = 35.6\%$ and $\eta = 1.55$ at S1 and $\rho = 56.3\%$ and
  $\eta = 2.29$ at S2, so the cumulative stages concentrate residual risk
  monotonically and not only improve the classification metrics.

  Containment holds at every stage and on every board, since the false-positive
  count is zero throughout. On the benchmark it held exactly: every one of the
  52 genuinely unmonitored signals remained in $A$. The in-sample residual
  error mode is thus redundant effort rather than escaped risk. A signal that
  escapes review is precisely a false positive: a genuine negative over-claimed as monitored and thereby removed
  from $A$. Because a monitored verdict is emitted only with a
  netlist-verified proof path, such an event can originate only from an
  incorrect admissible edge (e.g., a mis-supplied device class), never from the
  search itself. Its observed rate is zero.
\textcolor{red}{The approximate 95\% upper confidence bound
on the false-positive rate is 5.8\%, based on the 52 negatives
in the main 421-signal benchmark.}

Whereas the containment guarantee above holds the false-positive rate fixed within a single procedure, \textcolor{red}{the cross-model study on the SCM subset ($N=47$; 33 positives and 14 negatives) asks whether the observed zero-false-positive behavior persists across language models.} To test this, every model was given the identical
package, namely the same deterministic Union-Find and BFS engines, the same
netlist, and the same device specifications. Each model ran those engines while
extracting the device pin-roles and controller identifications itself, with no
model-specific tuning. Because the engines are identical across models, the
experiment isolates the model-dependent semantic extraction step from the
deterministic verifier, so any difference in outcome reflects how completely a
model extracts the semantics the engines consume, not the determinism of the
engines. In the SCM module, \textcolor{red}{Model A}, \textcolor{red}{Model B} and \textcolor{red}{Model C} were run five
times from a clean context and scored against the same expert ground truth.
Precision and specificity remained at 100\% in every run, while mean accuracy and
variance differed across models. These cross-model results are reported in
Table~\ref{tab:crossmodel}.

\begin{table}[!t]
  \centering
  \caption{Cross-model reproducibility of the NetFlow-DSR method.}
  \label{tab:crossmodel}
  \renewcommand{\arraystretch}{1.2}
  \setlength{\tabcolsep}{4pt}
  \resizebox{\columnwidth}{!}{%
  \begin{tabular}{@{}lccccc@{}}
  \toprule
  \textbf{Model} & \textbf{Acc \%} & \textbf{Sens \%} & \textbf{Prec \%} & \textbf{F1} & \textbf{Time}\textcolor{black}{\textsuperscript{*}} \\
  \midrule
  \textbf{\textcolor{red}{Model A}} & $\mathbf{88.52 \pm 1.17}$ & $\mathbf{83.63 \pm 1.65}$ & $\mathbf{100}$ & $\mathbf{0.911 \pm 0.010}$ &
  $\approx 6$ \\
  \textcolor{red}{Model B} & $88.08 \pm 2.85$ & $83.03 \pm 4.07$ & $100$ & $0.907 \pm 0.025$ & $\approx 3$ \\
  \textcolor{red}{Model C} & $80.43 \pm 8.16$ & $72.12 \pm 11.62$ & $100$ & $0.834 \pm 0.077$ & $\approx 21$ \\
  \bottomrule
  \end{tabular}%
  }
  \par\vspace{2pt}
  {\raggedright\footnotesize
  \textcolor{black}{\textsuperscript{*}Time is reported in minutes.}\par}
\end{table}
As summarized in Table~\ref{tab:crossmodel}, \textcolor{red}{Model A} and \textcolor{red}{Model B} achieved
comparable mean accuracy ($88.52\% \pm 1.17$ and $88.08\% \pm 2.85$,
respectively), with \textcolor{red}{Model A} showing the lowest variance and \textcolor{red}{Model B} the fastest
average runtime at approximately 3 minutes per run. \textcolor{red}{Model C} trailed at
$80.43\% \pm 8.16$ and was the slowest at approximately 21 minutes. Its lower
accuracy and higher variance reflect less complete and less consistent execution
of the procedure's semantic steps, not any loss of precision. The $\pm$ values are
standard deviations over five clean-slate runs. The run-to-run and
model-to-model differences come from how many device-mediated telemetry signals
each model extracted from the device specifications. Four of \textcolor{red}{Model B}'s five runs
reproduced the deterministic result around (89.36\% accuracy), while a single run
missed three sense-to-I2C signals (82.98\%), and no run over-claimed. The decisive
observation is a clean decomposition. Precision shows zero variance across models
($100\% \pm 0$ in every run), whereas accuracy and sensitivity variance grow only
with how completely a model extracts the device semantics. This is what the
architecture predicts, because the deterministic engines assert only
netlist-verified paths and therefore fix observed precision in the evaluated data,
while LLM-supplied semantics, namely device pin-role extraction, signal detection,
and controller identification, govern recall. Consistent with this decomposition,
the pin-role component that varied across models was the device class extracted
from the device specifications and its monitoring semantics, not pin
directionality, which is read deterministically from the design library and is
therefore identical for every model. Under correct device semantics, reliability thus follows from the netlist-grounded procedure rather than any single model. Given correct device semantics, the model determines how much is recovered, while precision depends on the verifier's inputs being correct.

  The main limitation of these results is the dependence on accurate extraction
 of device roles from device specifications. When
  design-library metadata is unavailable, connectivity can still be recovered
 from the netlist, but pin roles are then inferred from device
  specification documents rather than treated as a deterministic design-library
  input.

\section{Conclusion}\label{sec:conclusion}
This paper presented NetFlow-DSR, a netlist-grounded, neuro-symbolic procedure for
automated detection and verification of diagnostic-signal reachability. The method
formulates DSR as two graph relations, an identity relation over
series-passive-renamed fragments and a power-guarded reachability relation from
management-controller seeds. Both are computed by deterministic engines over the
netlist, while an LLM extracts device semantics from device specifications to
extend the graph only where component function justifies additional
observability. Across five anonymized production boards spanning 421 in-scope
signals, the cumulative procedure raises pooled recall from 40.7\% to 89.4\% and
F1 from 0.578 to 0.944, reaching up to 97.1\% recall on the PCIe board. Observed
precision and specificity hold at 100\%, with zero false positives in the
evaluated data. The role of structure and grounding, rather than any single
model, is made explicit by the cross-model study. When the same procedure was
executed on the SCM by \textcolor{red}{Model A}, \textcolor{red}{Model B}, and \textcolor{red}{Model C} over five clean-slate runs
each, precision remained exactly 100\% ($\pm 0$) for all three models even as mean
accuracy varied (88.5\%, 88.1\%, and 80.4\%, respectively) and wall-clock time
ranged from roughly 3 to 21 minutes. The property that matters most for a
serviceability sign-off, that no unmonitored signal in the evaluated data was
reported as covered, therefore tracks the structured, netlist-grounded procedure
rather than a particular language model. This zero-false-positive behavior is
structural rather than incidental. The verifier emits a monitored verdict only when
the admissibility predicate exhibits a netlist-verified proof path, so it is
sound by construction with respect to that model. A false positive can originate
only from an incorrect admissible edge, never from the search. Every
positive verdict carries a proof path that concentrates engineering effort on the
flagged unmonitored set rather than on repeated manual tracing. NetFlow-DSR thus
offers a scalable, auditable foundation for DSR automation in reliable electronic
systems. For datacenter platforms, where one design is replicated across a fleet,
it moves coverage gaps into the design review, before they are deployed at scale.
In this foundation, netlist grounding and explicit structure supply the reliability
and learning is reserved for the semantics a netlist alone cannot encode. Because the diagnostic scope is a configurable filter over a complete per-board
analysis, the same procedure serves narrow sign-off reviews and full-design audits
alike.

Several directions for future work follow from this study. First, widening the
declared controller scope to destinations that currently fall outside it, such as
CPU-hosted DIMM telemetry, would let scope-dependent signals be judged against the
correct endpoint. Second, joining per-board trace cards at shared connector pins would
extend the same proof path across a full tray, so each cross-board signal keeps one
identity card and system-level review reduces to checking the joined cards. Third,
coupling NetFlow-DSR's proof paths with firmware and sensor-configuration checks would close the gap between electrical reachability
and end-to-end diagnosability.

\bibliographystyle{IEEEtran}
\bibliography{reference_V43}

\end{document}